\documentclass[%
reprint,
superscriptaddress,
 amsmath,amssymb,
 aps,
floatfix,
]{revtex4-2}

\usepackage{graphicx,setspace}
\usepackage{dcolumn}
\usepackage{bm}
\usepackage[mathlines]{lineno}
\usepackage[colorlinks=true,citecolor=teal,linkcolor=olive,anchorcolor=green,urlcolor=teal]{hyperref}
\usepackage{mathrsfs}
\usepackage{float}
\usepackage{ulem}
\usepackage[utf8]{inputenc}
\usepackage{multirow}
 \usepackage[table,xcdraw]{xcolor}

\newcommand{\nn}{\nonumber}

\begin{document}

\title{Internal structures and circular orbits for test particles}
\author{Ming Zhang}
\email{mingzhang@jxnu.edu.cn}
\affiliation{Department of Physics, Jiangxi Normal University, Nanchang 330022, China}
\author{Jie Jiang}
\email{jiejiang@mail.bnu.edu.cn (corresponding author)}
\affiliation{College of Education for the Future, Beijing Normal University, Zhuhai 519087, China}
\date{\today}

\begin{abstract}
We explore how the internal structure of a test particle affects its equatorial stable circular orbits around the Kerr black hole with or without a cosmological constant. To this end, we first explicitly write equations of motion for a test particle in the pole-dipole-quadrupole approximation specifying the quadrupole momentum tensor to a spin-induced model. Then we calculate characteristic quantities -- radius, angular momentum, energy, angular velocity, and impact parameter -- for the  particles on the stable circular orbits. Once the pole-dipole-quadrupole approximation is taken, we find that  for a  particle on an innermost stable circular orbit, all characteristic quantities, except the angular velocity, become greater relative to the pole-dipole case. In contrast, for a  particle on an outermost stable circular orbit, which only exists in the case of the spacetime background being asymptotically de Sitter, it is the radius that becomes smaller while all other quantities become greater.
\end{abstract}

\maketitle

\section{Introduction}
The exploration of circular motion for a test particle in the strong gravitational field has been conducted ever since the emergence of General Relativity. It is well-known that the innermost stable circular orbit (ISCO) corresponds to the last stable orbit for a massive particle that orbits a central massive object, such as a neutron star or a black hole \cite{wald1984general}. The ISCO can be used to distinguish inspiralling and merging stages of a compact binary system with extreme mass ratio, where the gravitational waves emit \cite{Suzuki:1997by,Koekoek:2011mm,dAmbrosi:2014llh}. It also plays a key role in modeling the accretion disc as well as the radiation spectrum around the central massive object \cite{Shakura:1972te,Page:1974he}. 

The ISCO of a massive test particle around a spherically symmetric black hole is unique. For the Schwarzschild spacetime, the radius of the ISCO for a massive particle is $6 M$ ($M$ denotes mass of the black hole) \cite{chandrasekhar1985mathematical}. For the extreme Reissner-Nordström black hole, the radius of the ISCO becomes $4 M$ \cite{Pugliese:2010ps}. The ISCO for an axially symmetric rotating black hole splits into two branches: one is prograde orbit (co-rotating orbit) and the other is retrograde orbit (counter-rotating orbit). For the extreme Kerr black hole case, the radius of the prograde ISCO is $M$ and it is $9 M$ for the retrograde counterpart \cite{bardeen1972jm}. 

For the asymptotically flat and asymptotically Anti-de Sitter (AdS) spacetime, the ISCO is the only stable circular orbit for a massive particle. However, this is not the case for the de Sitter (dS) spacetime. As we know, the positive cosmological constant, though being very small ($\sim 10^{-52} \rm{m}^{-2}$) \cite{Spergel:2006hy}, is introduced to explain the accelerating expansion of the Universe at a large scale as well as the energy density of the vacuum \cite{Carroll:2000fy}. For the asymptotically dS spacetime with a positive cosmological constant, there are outermost stable circular orbit (OSCO) for massive particles besides the ISCO \cite{kottler:1918}. Investigations on massive particles' OSCOs around the Schwarzschild dS black hole can be found, for example, in Refs. \cite{howes:1979,Stuchlik:1999qk,Boonserm:2019nqq}.

To study the motion of a test particle, besides the self-force effect \cite{Barack:2009ey}, the finite-size effect cannot be neglected \cite{mathisson2010republication,Papapetrou:1951pa}. The internal structure of the particle can be modeled using the pole-dipole-quadrupole or higher multipole momentum structures \cite{Steinhoff:2009tk}. At least, the simplest case is to take into account the pole-dipole approximation for the particle. The ISCOs of the spinning particles in the pole-dipole approximation around the asymptotically flat or AdS black holes were studied extensively, for instance, in \cite{Harms:2016ctx,Lukes-Gerakopoulos:2017vkj,Zhang:2017nhl,Hojman:1976kn,Conde:2019juj,Toshmatov:2019bda}. In the background of the Kerr-dS spacetime, the ISCOs of the spinning particles in the pole-dipole approximation were investigated in \cite{Zhang:2018omr}. Then in \cite{Toshmatov:2020wky}, the OSCOs of the spinning particles were studied in the background of a Schwarzschild-dS black hole in the pole-dipole approximation. It was found that the OSCO, where the gravitational attraction of the black hole is balanced by the cosmological repulsive vacuum effect for a particle,  is sensitive to the variation of the cosmological constant.

We notice that it is still a question that how the higher multipole momenta of the test particles affect the particles' ISCOs and OSCOs. That is, how the higher multiple momenta modify the characteristic parameters of the ISCOs and OSCOs for the spinning particles. In this paper, we aim to study this question by considering the pole-dipole-quadrupole approximation of the test particle through specifying the quadrupole momentum tensor of the  particle to the spin-induced ``spin-squared” case. We will investigate how the pole-dipole-quadrupole approximation  modifies the circular orbit relative to the pole-dipole approximation  case for the particle. The rest of this paper will be arranged as follows. In Sec. \ref{sec:eos}, we will work out the equations of motion for the test particles in the Kerr spacetime with a cosmological constant. Subsequently, in Sec. \ref{sec:co}, we will investigate how the pole-dipole-quadrupole approximation of the particle affects its ISCOs and OSCOs relative to the case where only the pole-dipole  approximation of the spinning particle is considered. Conclusions will be drawn in Sec. \ref{sec:con}. In this paper, we name ISCO together with OSCO as marginally stable circular orbit (MSCO).

\section{Equations of motion for test particles}\label{sec:eos}

\subsection{Basics}
The spacetime background we consider is the Kerr black hole with a cosmological constant $\Lambda$, which in the Boyer-Lindquist coordinates reads
\begin{equation}\label{metric}
\begin{aligned}
\rm{d}s^2=&-\frac{\Delta}{\Sigma}\left(\frac{{\rm{d}} t}{\Xi}-\frac{a \sin ^{2} \theta}{\Xi} {\rm{d}} \phi\right)^{2}+\frac{\Sigma}{\Delta} {\rm{d}} r^{2}+\frac{\Sigma}{\Delta_{\theta}} {\rm{d}} \theta^{2}\\&+\frac{\Delta_{\theta} \sin ^{2} \theta}{\Sigma}\left(\frac{a {\rm{d}} t}{\Xi}-\frac{r^{2}+a^{2}}{\Xi} {\rm{d}} \phi\right)^{2},
\end{aligned}
\end{equation}
where
\begin{align}
\Sigma &=r^2+a^2 \cos ^2\theta,\nonumber~\\ \Delta &=\left(1-\frac{\Lambda r^2}{3}\right)\left(r^{2}+a^2\right)-2 Mr\nonumber,\\
\Xi &=1+\frac{\Lambda a^2}{3},\nonumber\\ \Delta_\theta &=1+\frac{\Lambda a^2}{3}\cos^2\theta.
\end{align}
$M$, $a$ are individually the mass and angular momentum per unit mass of the black hole. For simplicity, we set $M=1$.

In curved spacetime, due to the action of the Lorentz-like force produced by the particle's spin, the motion of the particle is no longer geodesic. To describe an extended test particle moving in the curved spacetime and consider the particle's multipole momentum up to the quadrupolar order, we use the Mathisson-Papapetrou-Dixon (MPD) equations truncated at the pole-dipole-quadrupole approximation \cite{Steinhoff:2009tk},
\begin{equation}\label{mpd1}
\frac{{\rm{D}} S^{\mu \nu}}{{\rm{D}} \tau}=2 p^{[\mu} u^{\nu]}-\frac{4}{3} R_{\alpha \beta \gamma}^{\,\,\,\quad[\mu} J^{\nu] \gamma \alpha \beta},
\end{equation}
\begin{equation}\label{mpd2}
\frac{{\rm{D}} p_{\mu}}{{\rm{D}} \tau}=-\frac{1}{2} R_{\mu \nu \alpha \beta} u^{\nu} S^{\alpha \beta}-\frac{1}{6} R_{\rho \alpha \beta \gamma ; \mu} J^{\rho \alpha \beta \gamma},
\end{equation}
where $R_{\mu\nu\alpha\beta}$ is the Riemann curvature tensor of the background spacetime, $\rm{D}/\rm{D}\tau=u^{\alpha} \nabla_{\alpha}$ is total derivative along the particle's world line with $\tau$ the proper time. $u^{\mu}\equiv {\rm{d}} x^{\mu} / {\rm{d}} \tau$ is the particle's four-velocity and it can be normalized to $u^{a}u_{a}=-1$ by a specific parametrization choice of the trajectories \cite{Obukhov:2010kn}. $S^{\mu\nu}$ is the particle's spin tensor, $p^\mu$ is the four-momentum of the particle. $J^{\mu\nu\alpha\beta}$ is the quadrupole tensor satisfying
\begin{equation}
J^{a b c d}=J^{[a b][c d]}=J^{c d a b},
\end{equation}
\begin{equation}
J^{[a b c] d}=0 \Leftrightarrow J^{a b c d}+J^{b c a d}+J^{c a b d}=0.
\end{equation}
The algebraic symmetries of the quadrupole tensor are the same as the Riemann tensor $R_{\mu\nu\alpha\beta}$. The dynamics of the particle cannot be sufficiently determined by the MPD equations and the normalization condition of the four-velocity. To close the MPD system of equations, we use the covariant Tulczyjew-Dixon spin supplementary condition (SSC) \cite{tulczyjew1959motion,dixon1964covariant,Dixon:1970zza}
 \begin{equation}\label{sscc}
 S^{\mu \nu} p_{\nu}=0,
 \end{equation}
 which is often used associated with extended test particles, so that three more equations are provided and the reference worldline defining the multipole momentum can be fixed.

In the following , we will consider test particles on the equatorial orbits \cite{Hojman:1976kn}, and set that the particle's spin is aligned with the black hole's spin or the spacetime rotation axis, so we have $S^\mu=-S\sqrt{\Delta_\theta}/\sqrt{\Sigma}\delta_\theta^\mu$, with $S$ the spin magnitude. Here $S>0$ leads to $S^\theta<0$, so that $S^\theta$ denotes a spin aligned to $\partial_z$ \cite{Steinhoff:2012rw}. The spin magnitude $S$ is related to the spin tensor and the spin four-vector by a relation 
\begin{equation}
S^{ab}S_{ab}=2S_a S^a=2 S^2.
\end{equation}
Then it immediately follows that
\begin{equation}\label{spmag}
\begin{aligned}
-S \frac{{{\rm{D}}} S}{{\rm{D}} \tau} &=-\frac{1}{2} \frac{{\rm{D}} S^{2}}{{\rm{D}} \tau}=\frac{1}{2} S_{a b} \frac{{\rm{D}} S^{a b}}{{\rm{D}}\tau} \\
&=S_{a b} p^{a} u^{b}-\frac{2}{3} S_{a b} R_{c d e}^{a} J^{b c d e}.
\end{aligned}
\end{equation}
We can see that the spin magnitude is conserved in the pole-dipole approximation with Tulczyjew-Dixon SCC in  aforementioned Eq. (\ref{sscc}). Generally, the spin magnitude is not conserved if the pole-dipole-quadrupole approximation is taken. However, we can see that it can be so with specific quadrupole momentum tensor, which will be introduced below.

From Eq. (\ref{mpd1}), we have 
\begin{equation}
p^\mu=-\frac{{\rm{D}} S^{\mu \nu}}{{\rm{D}} \tau}u_\nu+m_0 u^\mu-\frac{4}{3} R_{\alpha \beta \gamma}^{\,\,\,\quad[\mu} J^{\nu] \gamma \alpha \beta}u_\nu,
\end{equation}
where we have defined the rest mass of the particle $m_0\equiv -p^\nu u_\nu$. Subsequently, we can obtain
\begin{equation}\label{eq13}
\begin{aligned}
m^2&=\frac{{\rm{D}} S^{\mu \nu}}{{\rm{D}} \tau}u_\nu p_\mu+m_0^2+\frac{4}{3} R_{\alpha \beta \gamma}^{\,\,\,\quad[\mu} J^{\nu] \gamma \alpha \beta}u_\nu p_\mu,
\end{aligned}
\end{equation}
\begin{equation}\label{eqmpsr}
\begin{aligned}
\frac{{\rm{D}} m}{{\rm{D}} \tau}=&\frac{{\rm{D}} p_{\mu}}{{\rm{D}} \tau} \frac{p_\nu}{m m_0}\left[\frac{{\rm{D}} S^{\mu\nu}}{{\rm{D}} \tau}-\frac{4}{3} R_{\alpha\beta\gamma}^{\quad\,\,[\mu} J^{\nu] \gamma\alpha\beta}\right]\\&-\frac{m}{6 m_0} \frac{{\rm{D}} R_{\rho\alpha\beta\gamma}}{{\rm{D}} \tau} J^{\rho\alpha\beta\gamma},
\end{aligned}
\end{equation}
where we have defined the other mass parameter $m^2\equiv -p^\mu p_\mu$. It means that the mass parameter $m$ is no longer constant once the pole-dipole-quadrupole approximation of the test particle is considered, as also shown in \cite{Steinhoff:2009tk,Steinhoff:2012rw}. In fact, in the pole-dipole-quadrupole approximation, we have $p^\mu=m_0 u^\mu+\mathcal{O}(\epsilon^2)$, and $m^2=m_0^2+\mathcal{O}(\epsilon^3)$ \cite{Steinhoff:2012rw}, with $\epsilon\ll 1$ the scaling of the multipole momentum and each $\ell$ multipole denoted as $\mathcal{O}(\epsilon^\ell)$.

We specify the quadrupole momentum tensor to the spin-induced ``spin-squared” case \cite{Hinderer:2013uwa,Bini:2017pee},
\begin{equation}\label{ssqm}
J^{\alpha \beta \gamma \delta}=-\frac{3 m_0}{m^3} p^{[\alpha} Q^{\beta][\gamma} p^{\delta]},
\end{equation}
where the quadrupole tensor $Q_{\alpha\beta}$ is defined as 
\begin{equation}
Q_{\alpha \beta}=S_{\alpha \gamma} S_{\beta}^{\gamma}.
\end{equation}
$S_{\mu\nu}$ is related to a spacelike spin four-vector $S^\alpha$ by a relation
\begin{equation}
S^{\mu \nu}=-\frac{\epsilon^{\mu \nu \alpha \beta} S_{\alpha} p_{\beta}}{\sqrt{-p_{\gamma} p^{\gamma}}},
\end{equation}
where $\epsilon_{\mu \nu \alpha \beta}$ is the Levi-Civita tensor. With the spin-induced quadrupole momentum tensor Eq.  (\ref{ssqm}) and the  Tulczyjew-Dixon SCC Eq.  (\ref{sscc}), we can prove that 
\begin{equation}\label{vanis}
\mathrm{D}S/\mathrm{D}\tau=0,
\end{equation}
which means that the spin magnitude can be conserved in this condition \cite{Steinhoff:2012rw}. (We show the details of the proof in Appendix \ref{appa})

For the spacetime we consider, the equations of motion for the particle admit conserved energy $\mathcal{E}$ and angular momentum $\mathcal{J}$, 
\begin{equation}\label{cquan}
C_{\xi}=p_{\mu} \xi^{\mu}-\frac{1}{2} S^{\mu \nu} \nabla_{\nu} \xi_{\mu},
\end{equation}
corresponding to the Killing vectors $\xi=\xi_t, \xi_\phi$, respectively. Notice that these quantities are conserved to all multipole momentum orders \cite{ehlers1977dynamics}.

To facilitate our manipulation as well as choose an observer, we put the test particle into the orthogonal normalized Carter tetrad $e^{(a)}_\mu$ \cite{Carter:1968rr,Seme:1993}, with
\begin{align}
e_\mu^{(0)}&=\sqrt{\frac{\Delta }{\Sigma }} \left(\frac{{\rm{d}}t}{\Xi}-\frac{a \sin^{2}\theta}{\Xi}{ \rm{d}}\phi \right),\\e_\mu^{(1)}&=\sqrt{\frac{\Sigma }{\Delta }}{\rm{d}}r,\\e_\mu^{(2)}&= \sqrt{\frac{\Sigma}{\Delta_\theta} }{\rm{d}}\theta,\\e_\mu^{(3)}&=\sin\theta\sqrt{\frac{\Delta_\theta}{\Sigma }} \left(-\frac{a {\rm{d}}t}{\Xi}+\frac{a^2+r^2}{\Xi}{\rm{d}}\phi\right).
\end{align}
We will derive the equations of motion for the particle with the help of this tetrad. Note that with the tetrad we choose, the only nonvanishing component of the spin four-vector is $S^{(2)}=-S$. From now on, we rescale $S$ by $s=S/(M m)$ \cite{Steinhoff:2012rw}.

\subsection{Equations of motion for a spinning particle in a pole-dipole approximation}

For readers' convenience, let us first write the equations of motion for a spinning particle in the pole-dipole approximation. The relation between the four-velocity and the four-momentum is useful to obtain the equations of motion for the spinning particle. In the pole-dipole approximation, we have \cite{Hojman:1976kn,Steinhoff:2012rw}
\begin{equation}\label{uvrela}
u^{a}=v^{a}+\frac{2 S^{a b} v^{c} R_{b c d e} S^{d e}}{S^{b c} R_{b c d e} S^{d e}+4 m^{2}},
\end{equation}
where 
\begin{equation}\label{vmopm}
v^a\equiv m_0 p^a /m^2. 
\end{equation}
Detailed derivation of the equation can be seen, for instance, in Refs. \cite{Steinhoff:2009tk,Saijo:1998mn,Jefremov:2015gza,Obukhov:2010kn}. According to Eqs. (\ref{cquan}), we have
\begin{equation}
\begin{aligned}
e=&\frac{3 a r-\Lambda  r^3 s+3 s}{r^2 \left(a^2 \Lambda +3\right)}v^{(3)}\\&+\frac{ r \sqrt{a^2 \left(9-3 \Lambda r^2\right)-3 r \left(\Lambda r^3-3 r+6\right)}}{r^2 \left(a^2 \Lambda +3\right)}v^{(0)},
\end{aligned}
\end{equation}
\begin{equation}
\begin{aligned}
j=&\frac{3 a^2 r+a s \left(-\Lambda r^3+3 r+3\right)+3 r^3}{r^2 \left(a^2 \Lambda +3\right)}v^{(3)}\\&+\frac{a r \sqrt{a^2 \left(9-3 \Lambda r^2\right)-3 r \left(\Lambda r^3-3 r+6\right)}}{r^2 \left(a^2 \Lambda +3\right)}v^{(0)}\\&+\frac{r s \sqrt{a^2 \left(9-3 \Lambda r^2\right)-3 r \left(\Lambda r^3-3 r+6\right)}}{r^2 \left(a^2 \Lambda +3\right)}v^{(0)},
\end{aligned}
\end{equation}
where $e\equiv \mathcal{E}/m,\,j\equiv \mathcal{J}/m$ are the energy and angular momentum of the spinning particle per unit mass. The impact factor of the particle can be defined as $\chi=j/e.$

Then we can express the four-momentum of the spinning particle in terms of the constants of motion,
\begin{align}
v^{(0)}&=\frac{r\left(a^2 \Lambda +3\right) \left(3 m s \mathcal{W}_1-r^3 (\Lambda s \mathcal{W}_1+3 e)-3 a r \mathcal{W}_2\right)}{\sqrt{9r^2+3\mathcal{W}_3}\mathcal{W}_4},\label{eq20}\\
v^{(3)}&=\frac{r^2 \left(a^2 \Lambda +3\right) \mathcal{W}_2}{\mathcal{W}_4},\label{eq21}\\
v^{(1)}&=\sigma_r \sqrt{-1+(v^{(0)})^2 - (v^{(3)})^2}=\sigma_r \sqrt{O_d},\label{eq22}
\end{align}
where $\sigma_r=1 \,\,\rm{or} \,\,-1$ is for the outgoing or ingoing particle. We have defined the radial effective potential of the spinning particle in the pole-dipole approximation as $O_d$ and denoted
\begin{align}
\mathcal{W}_1&=ae-j,\nn\\
\mathcal{W}_2&=j-e (a+s),\nn\\
\mathcal{W}_3&=a^2 \left(3-\Lambda r^2\right)+3 r (r-2 m)-\Lambda r^4,\nn\\
\mathcal{W}_4&=r^3 \left(\Lambda s^2+3\right)-3 m s^2.\nn
\end{align}
Substituting the four-momentum into Eq. (\ref{uvrela}), we get the spinning particle's four-velocity as 
\begin{align}
u^{(0)}&=\left(1-\frac{9 m s^2 u^{(3)}}{\mathcal{W}_5-3 r^3}\right)v^{(0)},\\
u^{(1)}&=\left(1-\frac{9 m s^2 u^{(3)}}{\mathcal{W}_5-3 r^3}\right)v^{(1)},\label{proone}\\
u^{(3)}&=\left(1-\frac{9 m s^2 \left(1+\left(u^{(3)}\right)^2\right)}{\mathcal{W}_5-3 r^3}\right)v^{(3)},
\end{align}
where
\begin{equation}
\mathcal{W}_5=6 m s^2 u^{(3)}+s^2 (u^{(3)}+1) \left(3 m-\Lambda r^3\right)+\Lambda r^3 s^2 u^{(3)}\nn.
\end{equation}
Using the relation
\begin{equation}
u^{a}=\frac{{\rm{d}} t}{{\rm{d}} \tau}\left(\frac{\partial}{\partial t}\right)^{a}+\frac{{\rm{d}} r}{{\rm{d}} \tau}\left(\frac{\partial}{\partial r}\right)^{a}+\frac{{\rm{d}} \phi}{{\rm{d}} \tau}\left(\frac{\partial}{\partial \phi}\right)^{a},
\end{equation}
and $u^\mu=e^\mu_{(a)}u^{(a)}$, we obtain the equations of motion for the spinning particle as
\begin{align}
u^t&=\frac{\left(a^2 \Lambda +3\right) \left(a u^{(3)} \sqrt{3\mathcal{W}_5}+3 a^2 u^{(0)}+3 r^2 u^{(0)}\right)}{3 r \sqrt{9r^2+3\mathcal{W}_3}},\\
u^r&=\sqrt{\frac{\Delta}{\Sigma}}u^{(1)},\label{rtau}\\
u^\phi&=\frac{\left(a^2 \Lambda +3\right) \left(u^{(3)} \sqrt{\mathcal{W}_5}+\sqrt{3} a u^{(0)}\right)}{3 r \sqrt{\mathcal{W}_5}}.
\end{align} 
We see that in the pole-dipole approximation, the radial velocity is aligned with the radial momentum for the spinning particle.

\subsection{Equations of motion for a particle with a spin-induced quadrupole momentum}
As the constants of motion in Eq. (\ref{cquan}) are conserved at all higher multipole, we can know that the momenta $v^{(0)}$ and $v^{(3)}$ in the pole-dipole-quadrupole approximation are the same with the ones in Eqs. (\ref{eq20}) and (\ref{eq21}) for the pole-dipole case. But the radial momentum $v^{(1)}$ is different with the one shown in Eq. (\ref{eq22}). As pointed out in Ref. \cite{Steinhoff:2012rw}, we may introduce the scalings of the multipole momentum as
\begin{equation}
\begin{aligned}
&m_0\sim u^a \sim  \mathcal{O}(\epsilon^0),\\
&\frac{{\rm{D}} p_{a}}{{\rm{D}} \tau}\sim S^{ab} \sim \mathcal{O}(\epsilon^1),\\
&\frac{{\rm{D}} S^{ab}}{{\rm{D}} \tau}\sim J^{abcd}\sim \mathcal{O}(\epsilon^2).
\end{aligned}
\end{equation}
Therefore, we have
\begin{equation}
\frac{{\rm{D}} m}{{\rm{D}} \tau}=-\frac{m}{6 m_0} \frac{{\rm{D}} R_{\rho\alpha\beta\gamma}}{{\rm{D}} \tau} J^{\rho\alpha\beta\gamma}+\mathcal{O}(\epsilon^3)
\end{equation}
based on Eq. (\ref{eqmpsr}). It is suggested by Ref. \cite{Steinhoff:2012rw} that a approximately masslike constant quantity $\mu$ can be defined,
\begin{equation}\label{mumfs}
\mu\equiv m-\frac{1}{2} \mathcal{F}_{a b} S_{c}^{a} S^{c b},
\end{equation}
where $\mathcal{F}_{ab}\equiv - R_{a c b d} p^{c} p^{d}/m^2$,
which confirms
\begin{equation}
\frac{\rm{D} \mu}{\rm{D} \tau}=0+\mathcal{O}\left(\epsilon^{3}\right).
\end{equation}
Then by employing (\ref{mumfs}), we obtain
\begin{equation}
-\frac{m}{\mu}=-1 +\frac{m}{2m_0^2 \mu}R_{acbd} v^c v^d S_{e}^{a} S^{e b},
\end{equation}
where $v^a\equiv m_0 p^a /m^2+\mathcal{O}(\epsilon^2)$ is also used \cite{Steinhoff:2012rw}. Consequently, we have
\begin{equation}
v^a v_a=-1-\frac{1}{2m^2}R_{acbd} v^c v^d S_{e}^{a} S^{e b},
\end{equation}
where, being consistent with our approximation, $\mu$ is replaced by $m$ on the right side as higher orders of $s$ are neglected \cite{Steinhoff:2012rw}. This result is also used in Refs. \cite{Bini:2017pee,Hinderer:2013uwa}.

According to the last equation, we have 
\begin{equation}
\begin{aligned}
&-\left(v^{(0)}\right)^{2}+\left(v^{(1)}\right)^{2}+\left(v^{(3)}\right)^{2}\\&=-1+\frac{s^2 \left[3 \mathcal{W}^2_1+r^2\right]}{r^5}+\frac{a^4 \Lambda ^2 s^2 \mathcal{W}^2_1}{3 r^5}\\&\quad-\frac{\Lambda s^2 \left[-6 a^4 e^2+6 a^2 j \left(2 a e-j\right)+r^5\right]}{3 r^5},
\end{aligned}
\end{equation}
from which the radial momentum in the tetrad can be obtained,
\begin{equation}\label{effq}
v^{(1)}=\sigma_r \sqrt{O_q}.
\end{equation}
The effective potential for the particle in the pole-dipole-quadrupole approximation is related to the one for the spinning particle in the pole-dipole approximation by
\begin{equation}
\begin{aligned}
O_q=&O_d +\frac{s^2 \left[3 \mathcal{W}^2_1+r^2\right]}{r^5}+\frac{a^4 \Lambda ^2 s^2 \mathcal{W}^2_1}{3 r^5}\\&-\frac{\Lambda s^2 \left[-6 a^4 e^2+6 a^2 j \left(2 a e-j\right)+r^5\right]}{3 r^5}.
\end{aligned}
\end{equation}

According to Eqs. (\ref{mpd2}) and (\ref{eq13}), we have (see Appendix \ref{uvqr} for detailed deduction)
\begin{equation}\label{eq30}
\begin{aligned}
u^{a}=& v^{a}\left(1+\frac{1}{2} Q^{b c} E_{b c}\right)+\frac{1}{2} S^{a b} R_{b c d f} v^{c} S^{d f} \\
&- R_{c d f}^{a} Q^{f d} v^{c},
\end{aligned}
\end{equation}
where $E_{b d}\equiv R_{a b c d} v^{a} v^{c}$.
According to Eq. (\ref{eq30}), if we do not set $p^{(1)}=0$, but only set $p^{(2)}=0$, we can get
\begin{equation}\label{equvge}
\begin{aligned}
u^{(a)}=&v^{(a)}\left[1-\frac{\left[\left(p^{(1)}\right)^2-1\right]\left[\left(p^{(1)}\right)^2+3 \left(p^{(3)}\right)^2+1\right] s^2}{2 r^3}\right.\\&\left.-\frac{\Lambda \left[\left(p^{(1)}\right)^4-1\right]s^2}{3}\right].
\end{aligned}
\end{equation}
Then we know that, like the pole-dipole case, the radial velocity is proportional to the radial component of the four-momentum even if the quadrupole momentum of the particle is included.

To the end of investigating the circular motion of the  particle on the equatorial plane, we can set $p^{(1)}=p^{(2)}=0$. According to Eq. (\ref{eq30}), we can get the relation between the four-velocity and the four-momentum in tetrad,
\begin{equation}\label{eq35}
u^{(a)}=v^{(a)}\left[1+\frac{\left(3 \left(p^{(3)}\right)^2+1\right) s^2}{2 r^3}+\frac{\Lambda s^2}{3}\right],
\end{equation}
from which we can obtain the four-velocity in the tetrad by using the four-momentum in the tetrad. We can further calculate the four velocity, which in the original coordinate system can be explicitly written as
\begin{equation}
\begin{aligned}
u^{t}=&\frac{ \left(a^2 \Lambda +3\right) \left(a^2+r^2\right)}{r \sqrt{a^2 \left(9-3 \Lambda r^2\right)-3 r \left(6 m+\Lambda r^3-3 r\right)}} u^{(0)}\\&+\frac{a \left(a^2 \Lambda +3\right)}{3 r} u^{(3)},
\end{aligned}
\end{equation}
\begin{equation}
\begin{aligned}
u^{\phi}=&\frac{a \left(a^2 \Lambda +3\right)}{r \sqrt{a^2 \left(9-3 \Lambda r^2\right)-3 r \left(6 m+\Lambda r^3-3 r\right)}}u^{(0)}\\&+\frac{(a^2 \Lambda +3)}{3 r}u^{(3)}.
\end{aligned}
\end{equation}

With the equations of motion obtained, we can explore how the internal structures influence the circular orbits of the  particles. Generally, the orbital parameters of a  particle include the radius, angular momentum, energy, impact parameter, and angular velocity. The angular velocity is defined as
\begin{equation}
\omega=\frac{u^\phi}{u^t}.
\end{equation}
In what follows, we will focus on these orbital parameters of the particle moving on an equatorial circular orbit.

It should be noted that there are restrictions to the parameters of the particles. Firstly, the forward-in-time condition $u^t>0$ should be satisfied \cite{Grib:2013hxa}. Secondly, the particle's spin must be small, $s\ll 1$, so that the backreaction effect of the  particle can be neglected. We will comply with these conditions when choosing different values of parameters in what follows. To ensure the existence of the event horizon and the cosmological horizon of the Kerr-dS black hole, the parameters of the black hole should also be restricted. Even all the conditions referred to are satisfied, MSCOs can still be lacking. The ranges of the parameters, especially for the cosmological constant, to ensure the existence of the MSCOs for the Kerr-dS black hole were elaborately discussed in \cite{Stuchlik:2003dt}.

\section{ISCO\lowercase{s} and OSCO\lowercase{s} of the particles}\label{sec:co}
Based on the result in Eq. (\ref{equvge}), we can use the radial effective potential $O_q$ in Eq. (\ref{effq}) to calculate the characteristic quantities of the MSCOs. The conditions which should be satisfied by the particles on the MSCOs are

\begin{table*}[]
\caption{Listed are characteristic quantities, $r,\,\omega,\,j,\,e,\,\chi$, for the test particles on MSCOs with $a=0.9, \Lambda=5\times 10^{-5}$. The second column is for the nonspinning particle, and variations of  characteristic quantities in the pole-dipole approximation as well as in the pole-dipole-quadrupole approximation for different spin parameters are shown in the third column and fourth column, respectively. $\kappa$ denotes the difference between the characteristic quantities' absolute value in the pole-dipole-quadrupole approximation and their counterparts  in the pole-dipole approximation. When the difference is positive, we denote it using the sign "+"; for a negative difference, we use the sign "--".}
\begin{tabular}{ccccccccccc}
\multicolumn{1}{c|}{} & \multicolumn{1}{c|}{}    & \multicolumn{4}{c|}{pole-dipole case}                                               & \multicolumn{4}{c|}{pole-dipole-quadrupole case}                                               &  \\ \hline
\multicolumn{1}{c|}{$s$}  & \multicolumn{1}{c|}{0}       & -0.1       & -0.05      & 0.05        & \multicolumn{1}{c|}{0.1}        & -0.1       & -0.05      & 0.05        & \multicolumn{1}{c|}{0.1}        & $\kappa$     \\ \hline
\multicolumn{1}{c|}{$r_{\rm{pi}}$} & \multicolumn{1}{c|}{2.3216}  & 2.4303     & 2.3757     & 2.2683      & \multicolumn{1}{c|}{2.2161}     & 2.4331     & 2.3764     & 2.2688      & \multicolumn{1}{c|}{2.2180}     & +   \\
\multicolumn{1}{c|}{$r_{\rm{ri}}$} & \multicolumn{1}{c|}{8.8869}  & 8.6918     & 8.7903     & 8.9816      & \multicolumn{1}{c|}{9.0745}     & 8.6960     & 8.7913     & 8.9825      & \multicolumn{1}{c|}{9.0780}     & +   \\
\multicolumn{1}{c|}{$r_{\rm{ro}}$} & \multicolumn{1}{c|}{23.1181} & 23.1590    & 23.1386    & 23.0975     & \multicolumn{1}{c|}{23.0768}    & 23.1587    & 23.1385    & 23.0974     & \multicolumn{1}{c|}{23.0765}    & --   \\
\multicolumn{1}{c|}{$r_{\rm{po}}$} & \multicolumn{1}{c|}{24.0974} & 24.0797    & 24.0886    & 24.10621    & \multicolumn{1}{c|}{24.11503}   & 24.0796    & 24.0885    & 24.10619    & \multicolumn{1}{c|}{24.11496}   & --   \\
\multicolumn{1}{c|}{$\omega_{\rm{pi}}$} & \multicolumn{1}{c|}{0.2253}  & 0.2163     & 0.2207     & 0.2301      & \multicolumn{1}{c|}{0.2350}     & 0.2160     & 0.2265     & 0.2300      & \multicolumn{1}{c|}{0.2347}     & --   \\
\multicolumn{1}{c|}{$\omega_{\rm{ri}}$} & \multicolumn{1}{c|}{-0.0388} & -0.03990   & -0.0394    & -0.03834    & \multicolumn{1}{c|}{-0.0379}    & -0.03987   & -0.0393    & -0.03833    & \multicolumn{1}{c|}{-0.0378}    & --   \\
\multicolumn{1}{c|}{$\omega_{\rm{ro}}$} & \multicolumn{1}{c|}{-0.0080} & -0.0080331 & -0.0080541 & -0.00809630 & \multicolumn{1}{c|}{-0.0081176} & -0.0080334 & -0.0080542 & -0.00809637 & \multicolumn{1}{c|}{-0.0081179} & +   \\
\multicolumn{1}{c|}{$\omega_{\rm{po}}$} & \multicolumn{1}{c|}{0.0074}  & 0.0073724  & 0.00736299 & 0.00734416  & \multicolumn{1}{c|}{0.0073348}  & 0.0073726  & 0.0073630  & 0.0073442   & \multicolumn{1}{c|}{0.0073349}  & +   \\
\multicolumn{1}{c|}{$j_{\rm{pi}}$} & \multicolumn{1}{c|}{2.0995}  & 2.0962     & 2.0989     & 2.0979      & \multicolumn{1}{c|}{2.0941}     & 2.0987     & 2.0996     & 2.0986      & \multicolumn{1}{c|}{2.0969}     & +   \\
\multicolumn{1}{c|}{$j_{\rm{ri}}$} & \multicolumn{1}{c|}{-4.1460} & -4.2040    & -4.1752    & -4.1163     & \multicolumn{1}{c|}{-4.0862}    & -4.2048    & -4.1754    & -4.1165     & \multicolumn{1}{c|}{-4.0868}    & +   \\
\multicolumn{1}{c|}{$j_{\rm{ro}}$} & \multicolumn{1}{c|}{-4.7313} & -4.8194    & -4.77535   & -4.68731    & \multicolumn{1}{c|}{-4.64328}   & -4.8195    & -4.77538   & -4.68733    & \multicolumn{1}{c|}{-4.64338}   & +   \\
\multicolumn{1}{c|}{$j_{\rm{po}}$} & \multicolumn{1}{c|}{4.4773}  & 4.3857     & 4.43150    & 4.52309     & \multicolumn{1}{c|}{4.5689}     & 4.3858     & 4.43152    & 4.52311     & \multicolumn{1}{c|}{4.5690}     & +   \\
\multicolumn{1}{c|}{$e_{\rm{pi}}$} & \multicolumn{1}{c|}{0.8441}  & 0.8540     & 0.8492     & 0.8385      & \multicolumn{1}{c|}{0.8325}     & 0.8543     & 0.8493     & 0.8386      & \multicolumn{1}{c|}{0.8328}     & +   \\
\multicolumn{1}{c|}{$e_{\rm{ri}}$} & \multicolumn{1}{c|}{0.9592}  & 0.95850    & 0.958879   & 0.959584    & \multicolumn{1}{c|}{0.95992}    & 0.95852    & 0.958883   & 0.959588    & \multicolumn{1}{c|}{0.95993}    & +   \\
\multicolumn{1}{c|}{$e_{\rm{ro}}$} & \multicolumn{1}{c|}{0.9700}  & 0.9699686  & 0.9699865  & 0.9700225   & \multicolumn{1}{c|}{0.9700406}  & 0.9699690  & 0.9699866  & 0.9700226   & \multicolumn{1}{c|}{0.9700410}  & +   \\
\multicolumn{1}{c|}{$e_{\rm{po}}$} & \multicolumn{1}{c|}{0.9694}  & 0.9694623  & 0.96945241 & 0.96943265  & \multicolumn{1}{c|}{0.9694228}  & 0.9694626  & 0.96945248 & 0.96943272  & \multicolumn{1}{c|}{0.9694231}  & +   \\
\multicolumn{1}{c|}{$\chi_{\rm{pi}}$} & \multicolumn{1}{c|}{2.4873}  & 2.4546     & 2.4715     & 2.5019      & \multicolumn{1}{c|}{2.5154}     & 2.4567     & 2.4721     & 2.5025      & \multicolumn{1}{c|}{2.5177}     & +   \\
\multicolumn{1}{c|}{$\chi_{\rm{ri}}$} & \multicolumn{1}{c|}{-4.3222} & -4.3860    & -4.3543    & -4.2897     & \multicolumn{1}{c|}{-4.2568}    & -4.3867    & -4.3545    & -4.2898     & \multicolumn{1}{c|}{-4.2574}    & +   \\
\multicolumn{1}{c|}{$\chi_{\rm{ro}}$} & \multicolumn{1}{c|}{-4.8776} & -4.9686    & -4.92311   & -4.83216    & \multicolumn{1}{c|}{-4.7867}    & -4.9687    & -4.92314   & -4.83219    & \multicolumn{1}{c|}{-4.7868}    & +   \\
\multicolumn{1}{c|}{$\chi_{\rm{po}}$} & \multicolumn{1}{c|}{4.6184}  & 4.523856   & 4.5711     & 4.66571     & \multicolumn{1}{c|}{4.7130}     & 4.523930   & 4.5712     & 4.66573     & \multicolumn{1}{c|}{4.7131}     & +   \\                       
\end{tabular}
\label{tab1}
\end{table*}

\begin{table*}[]
\caption{Listed are characteristic quantities, $r,\,\omega,\,j,\,e,\,\chi$, for the test particles on ISCOs with $a=0.9, \Lambda=0$.}
\begin{tabular}{c|c|cccc|cccc|c}
                      &                          & \multicolumn{4}{c|}{pole-dipole case}                                       & \multicolumn{4}{c|}{pole-dipole-quadrupole case}                   &    \\ \hline
\multicolumn{1}{c|}{$s$}  & \multicolumn{1}{c|}{0}       & \multicolumn{1}{c}{-0.1}     & -0.05     & 0.05      & 0.1      & -0.1     & -0.05     & 0.05      & 0.1      & $\kappa$    \\ \hline
\multicolumn{1}{c|}{$r_{\rm{pi}}$} & \multicolumn{1}{c|}{2.3209}  & \multicolumn{1}{c}{2.4294}   & 2.3749    & 2.2676    & 2.2155   & 2.4322   & 2.3756    & 2.2682    & 2.2174   & +  \\
\multicolumn{1}{c|}{$r_{\rm{ri}}$} & \multicolumn{1}{c|}{8.7174}  & \multicolumn{1}{c}{8.5365}   & 8.6280    & 8.8048    & 8.8903   & 8.5404   & 8.6289    & 8.8056    & 8.8936   & +  \\
\multicolumn{1}{c|}{$\omega_{\rm{pi}}$} & \multicolumn{1}{c|}{0.2254}  & \multicolumn{1}{c}{0.2164}   & 0.22084   & 0.23019   & 0.2351   & 0.2163   & 0.22080   & 0.23017   & 0.2350   & -- \\
\multicolumn{1}{c|}{$\omega_{\rm{ri}}$} & \multicolumn{1}{c|}{-0.0403} & \multicolumn{1}{c}{-0.04125} & -0.040748 & -0.039795 & -0.03935 & -0.04124 & -0.040742 & -0.039791 & -0.03933 & -- \\
\multicolumn{1}{c|}{$j_{\rm{pi}}$} & \multicolumn{1}{c|}{2.0998}  & \multicolumn{1}{c}{2.0965}   & 2.0992    & 2.0982    & 2.0943   & 2.0991   & 2.0999    & 2.0989    & 2.0971   & +  \\
$j_{\rm{ri}}$                      & -4.1681                      & -4.2248                      & -4.1967   & -4.1390   & -4.1095  & -4.2256  & -4.1969   & -4.1392   & -4.1102  & +  \\
$e_{\rm{pi}}$                      & 0.8442                       & 0.8542                       & 0.8494    & 0.8387    & 0.8327   & 0.8545   & 0.8495    & 0.8388    & 0.8330   & +  \\
$e_{\rm{ri}}$                      & 0.9610                       & 0.96021                      & 0.96061   & 0.96137   & 0.96173  & 0.96023  & 0.96062   & 0.96138   & 0.96175  & +  \\
$\chi_{\rm{pi}}$                      & 2.4872                       & 2.4545                       & 2.4714    & 2.5018    & 2.5152   & 2.4566   & 2.4720    & 2.5024    & 2.5176   & +  \\
$\chi_{\rm{ri}}$                      & -4.3372                      & -4.3999                      & -4.3687   & -4.3053   & -4.2731  & -4.4006  & -4.3689   & -4.3055   & -4.2737  & + 
\end{tabular}
\label{tab2}
\end{table*}

\begin{table*}[]
\caption{Listed are characteristic quantities, $r,\,\omega,\,j,\,e,\,\chi$, for the test particles on ISCOs with $a=0.9, \Lambda=-5\times 10^{-5}$.}
\begin{tabular}{c|c|cccc|cccc|c}
&     & \multicolumn{4}{c|}{pole-dipole case}                   & \multicolumn{4}{c|}{pole-dipole-quadrupole case}                   &   \\ \hline
$s$  & 0       & -0.1     & -0.05     & 0.05      & 0.1      & -0.1     & -0.05     & 0.05      & 0.1      &   $\kappa$  \\ \hline
$r_{\rm{pi}}$ & 2.3202  & 2.4285   & 2.3741    & 2.2670    & 2.2149   & 2.4312   & 2.3748    & 2.2675    & 2.2168   & +  \\
$r_{\rm{ri}}$ & 8.5774  & 8.4067   & 8.4931    & 8.6596    & 8.7400   & 8.4105   & 8.4940    & 8.6605    & 8.7432   & +  \\
$\omega_{\rm{pi}}$ & 0.2255  & 0.2165   & 0.22095   & 0.23029   & 0.2352   & 0.2164   & 0.22091   & 0.23026   & 0.2351   & -- \\
$\omega_{\rm{ri}}$ & -0.0415 & -0.04246 & -0.041976 & -0.041069 & -0.04065 & -0.04244 & -0.041971 & -0.041064 & -0.04063 & -- \\
$j_{\rm{pi}}$ & 2.1001  & 2.097    & 2.0995    & 2.098     & 2.095    & 2.099    & 2.1002    & 2.099     & 2.097    & +  \\
$j_{\rm{ri}}$ & -4.1889 & -4.244   & -4.2169   & -4.1604   & -4.131   & -4.245   & -4.2171   & -4.1606   & -4.132   & +  \\
$e_{\rm{pi}}$ & 0.8444  & 0.8543   & 0.84957   & 0.8388    & 0.8328   & 0.8546   & 0.84964   & 0.8389    & 0.8331   & +  \\
$e_{\rm{ri}}$ & 0.9627  & 0.96188  & 0.962307  & 0.963116  & 0.96350  & 0.96190  & 0.962311  & 0.963121  & 0.96352  & +  \\
$\chi_{\rm{pi}}$ & 2.4870  & 2.4544   & 2.4713    & 2.5016    & 2.5151   & 2.4566   & 2.4719    & 2.5022    & 2.5174   & +  \\
$\chi_{\rm{ri}}$ & -4.3511 & -4.4127  & -4.3821   & -4.3197   & -4.2880  & -4.4135  & -4.3823   & -4.3199   & -4.2886  & + 
\end{tabular}
\label{tab3}
\end{table*}

\begin{align}
O_q&\equiv\mathcal{F}=0,\label{conone}\\
\frac{{\rm{d}} O_q}{{\rm{d}}{r}}&\equiv\mathcal{G}=0,\label{contwo}\\
\frac{{\rm{d}}^2 O_q}{{\rm{d}}r^2}&\equiv\mathcal{H}=0,\label{conthree}
\end{align}
where 
\begin{equation}
\begin{aligned}
O_q=&\frac{r^2 \left(a^2 \Lambda +3\right)^2 \left[3 a r W_2-3 e r^3+W_1 \left(\Lambda r^3 s-3 m s\right)\right]^2}{3 W_3 W_4^2}\\&-\frac{r^4 W_2^2 \left(a^2 \Lambda +3\right)^2}{W_4^2}-1+\frac{s^2 \left[3 \mathcal{W}^2_1+r^2\right]}{r^5}\\&+\frac{a^4 \Lambda ^2 s^2 \mathcal{W}^2_1}{3 r^5}-\frac{\Lambda s^2 \left[6 a^2 j \left(2 a e-j\right)+r^5-6 a^4 e^2\right]}{3 r^5}.
\end{aligned}
\end{equation}
Eq. (\ref{conone}) corresponds to the turning points of the test particle on the equatorial plane. Eqs. (\ref{conone}) and (\ref{contwo}) provide circular equatorial orbits. The last one Eq. (\ref{conthree}) corresponds to the marginally stable condition \cite{Boonserm:2019nqq,Favata:2010ic}. To solve the equations (\ref{conone}), (\ref{contwo}) and (\ref{conthree}), we refer to the resultant method \cite{Chakraborty:2013kza,Hod:2013qaa,Zaslavskii:2014mqa,Toshmatov:2020wky,Schroven:2021zux}. To know about the details of the resultant method as well as its application, one can refer to Refs. \cite{cox2013ideals,Beheshti:2015bak}.

We now denote $\mathcal{F},\,\mathcal{G}\,,\mathcal{H}$ in Eqs. (\ref{conone}), (\ref{contwo}) and (\ref{conthree}) as 
\begin{align}
\mathcal{F}&=a_{0} x^{2}+a_{1} x y+a_{2} y^{2}+a_{3} x z+a_{4} y z+a_{5} z^{2},\\
\mathcal{G}&=b_{0} x^{2}+b_{1} x y+b_{2} y^{2}+b_{3} x z+b_{4} y z+b_{5} z^{2},\\
\mathcal{H}&=c_{0} x^{2}+c_{1} x y+c_{2} y^{2}+c_{3} x z+c_{4} y z+c_{5} z^{2},
\end{align}
where $x\equiv e\,,y\equiv j\,$ and the parameter $z$ is introduced by rescaling $e$ and $j$ through $x \mapsto x / z, y \mapsto y / z$. We have $a_3=a_4=b_3=b_4=b_5=c_3=c_4=c_5=0$, and the specific expressions of other $a_i,\,b_i\,,c_i$ are too tedious to show here, which can be obtained by simple algebraic calculations. (The same for what follows.) Then we get a Jacobian determinant $J$, which is defined by 
\begin{equation}
J=J(x, y, z)\equiv\operatorname{det}\left(\begin{array}{lll}
F_{x} & F_{y} & F_{z} \\
G_{x} & G_{y} & G_{z} \\
H_{x} & H_{y} & H_{z}
\end{array}\right),
\end{equation}
where the subscripts mean partial derivatives. Thus, we have 
\begin{align}
J_{x}&=\zeta_{0} x^{2}+\zeta_{1} x y+\zeta_{2} y^{2}+\zeta_{3} x z+\zeta_{4} y z+\zeta_{5} z^{2},\\
J_{y}&=\iota_{0} x^{2}+\iota_{1} x y+\iota_{2} y^{2}+\iota_{3} x z+\iota_{4} y z+\iota_{5} z^{2},\\
J_{z}&=\varpi_{0} x^{2}+\varpi_{1} x y+\varpi_{2} y^{2}+\varpi_{3} x z+\varpi_{4} y z+\varpi_{5} z^{2},
\end{align}
where $\zeta_{0}=\zeta_{1}=\zeta_{2}=\zeta_{5}=\iota_{0}=\iota_{1}=\iota_{2}=\iota_{5}=\varpi_{3}=\varpi_{4}=\varpi_{5}=0.$ Lastly, we can obtain the resultant of $\mathcal{F}$, $\mathcal{G}$ and $\mathcal{H}$,
\begin{equation}
\operatorname{Res}(\mathcal{F}, \mathcal{G}, \mathcal{H})\equiv\operatorname{det}\left(\begin{array}{llllll}
a_{0} & b_{0} & c_{0} & \zeta_{0} & \iota_{0} & \varpi_{0} \\
a_{1} & b_{1} & c_{1} & \zeta_{1} & \iota_{1} & \varpi_{1} \\
a_{2} & b_{2} & c_{2} & \zeta_{2} & \iota_{2} & \varpi_{2} \\
a_{3} & b_{3} & c_{3} & \zeta_{3} & \iota_{3} & \varpi_{3} \\
a_{4} & b_{4} & c_{4} & \zeta_{4} & \iota_{4} & \varpi_{4} \\
a_{5} & b_{5} & c_{5} & \zeta_{5} & \iota_{5} & \varpi_{5}
\end{array}\right).
\end{equation}
The radii of the ISCOs and the OSCOs can be obtained by solving the equation
\begin{equation}\label{resultant1}
\operatorname{Res}(\mathcal{F}, \mathcal{G}, \mathcal{H})=0.
\end{equation}

With the radii of the MSCOs, we then can calculate other characteristic quantities of the  particles on the MSCOs by employing any two of Eqs. (\ref{conone}), (\ref{contwo}) and (\ref{conthree}). In fact, it is not easy to get the characteristic quantities directly only by root-finding approach from Eqs. (\ref{conone}), (\ref{contwo}) and (\ref{conthree}), especially once the terms containing cosmological constant are included. However, we can do from simple sanity checks that the result obtained from the resultant method definitely solves Eqs. (\ref{conone}), (\ref{contwo}) and (\ref{conthree}).

Focusing on how the internal structure of the  particle affects its MSCOs, we will compare the characteristic quantities obtained in the pole-dipole approximation with the ones in the pole-dipole-quadrupole approximation. For the Kerr-dS case, we find both ISCOs and OSCOs for the particles. To denote the radius, conserved angular momentum, conserved energy, angular velocity, and the impact factor for the  particle, we use $r_{\rm{pi}}$, $j_{\rm{pi}}$, $e_{\rm{pi}}$, $\omega_{\rm{pi}}$, $\chi_{\rm{pi}}$ for the prograde ones on the ISCO, $r_{\rm{ri}}$, $j_{\rm{ri}}$, $e_{\rm{ri}}$, $\omega_{\rm{ri}}$, $\chi_{\rm{ri}}$ for the retrograde ones on the ISCO, $r_{\rm{po}}$, $j_{\rm{po}}$, $e_{\rm{po}}$, $\omega_{\rm{po}}$, $\chi_{\rm{po}}$ for the prograde ones on the OSCO, and $r_{\rm{ro}}$, $j_{\rm{ro}}$, $e_{\rm{ro}}$, $\omega_{\rm{ro}}$, $\chi_{\rm{ro}}$ for the retrograde ones on the OSCO. To show how the internal structures of the  particle affect its characteristic quantities, we will compare the values of the characteristic quantities in the pole-dipole approximation and the ones in the pole-dipole-quadrupole approximation, respectively.

We list our calculation results of characteristic quantities for the Kerr-dS case in Table \ref{tab1}, where both the quantities for the particles on the  ISCOs and the ones on the OSCOs exist. Also, we calculate the characteristic quantities for the Kerr case and the Kerr-AdS case, where only ISCOs exist, and list related results in Tables \ref{tab2} and \ref{tab3}, respectively. From the Tables \ref{tab1}, \ref{tab2} and \ref{tab3}, we can see that: (1) For a  particle on the ISCO in the Kerr spacetime with or without a cosmological constant, all its characteristic quantities, except the angular velocity, become greater in the pole-dipole-quadrupole approximation, comparing with their counterparts in the pole-dipole approximation; (2) For a  particle on the OSCO in the Kerr-dS spacetime, all its characteristic quantities, except the radius, become greater in the pole-dipole-quadrupole approximation, comparing with their counterparts in the pole-dipole approximation.

Notice that though we only present several tables with separate numeric values, we have checked that the characters of the shifts of the characteristics due to the quadrupole terms are respected for all cases in the parameter space.

On the other hand, we can do some comparisons with the results obtained in existing literature. In Ref. \cite{Jefremov:2015gza}, the circular motion of the spinning particle in the pole-dipole approximation was studied in the Kerr background. It is easy to find that the changing tendencies of the characteristic quantities we obtained in Table \ref{tab2} is the same with those shown in the explicit expressions for special cases such as case of extreme Kerr spacetime background. Beside, we reproduced the results for the spinless particles or spinning particles in the pole-dipole approximation with a dS spacetime background, which are investigated in Ref.  \cite{Boonserm:2019nqq} and Refs. \cite{Zhang:2018omr,Toshmatov:2020wky}, respectively. What makes a difference here is that we further calculated the ISCOs and OSCOs in the pole-dipole-quadrupole approximation.

\section{Conclusions}\label{sec:con}
We studied the effects of the  particle's internal structures on the particle's MSCOs. In the spacetime background of the Kerr black hole with (without) a cosmological constant, we first wrote the equations of motion for the spinning particle in the pole-dipole approximation and then further derived those equations for the particle in the pole-dipole-quadrupole approximation. We investigated the ISCOs and OSCOs of the  particles around the Kerr-dS black hole, and also explored the ISCOs for the Kerr(-AdS) black hole case by calculating characteristic quantities, radius, angular momentum, energy, angular velocity and impact parameter, of the particles on the MSCOs. On the one hand, for the  particles on the ISCOs, regardless of the spacetime's asymptotical structure, our results showed that the values of the characteristic radius, angular momentum, energy, and impact parameter obtained in the pole-dipole approximation are always smaller than the ones obtained in the pole-dipole-quadrupole approximation. However, the characteristic angular velocity is exceptional. On the other hand, for the case of the  particles locating on the OSCOs, which only exists for the asymptotically dS spacetime, we found that the values of the characteristic angular velocity, angular momentum, energy, and impact parameter obtained in the pole-dipole approximation are always smaller than the ones obtained in the pole-dipole-quadrupole approximation. However, the characteristic radius is exceptional.

We should point out that the equations of motion Eqs. (\ref{mpd1}) and (\ref{mpd2}) are schemes approximated to the pole-dipole-quadrupole level, capturing certain features of the motion for the  extended particles. Our theoretical results may benefit the study of both single black hole physics and binary black hole systems. To develop the present study, a completely general quadrupole momentum including gravito-electric tidal field and gravito-magnetic tidal field should be encoded in the MPD equation \cite{Steinhoff:2012rw}. We leave it for future work.

\section*{Acknowledgements}
M. Z. is supported by the National Natural Science Foundation of China with Grant No. 12005080.  J. J. is supported by the National Natural Science Foundation of China with Grant No. 12205014, the Guangdong Basic and Applied Research Foundation with Grant No. 217200003 and the Talents Introduction Foundation of Beijing Normal University with Grant No. 310432102.

\appendix

\section{Derivation of Eq. (\ref{vanis})}\label{appa}
According to Eqs. (\ref{spmag}) and (\ref{ssqm}), we have
\begin{equation}\begin{aligned}
-S \frac{{{\rm{D}}} S}{{\rm{D}} \tau} &=S_{a b} p^{a} u^{b}+\frac{2m_0}{m^3} S_{a b} R_{c d e}^{a} p^{[b}S^{c]f}S_f^{[d}p^{e]}\\&=S_{a b} p^{a} u^{b}+\frac{m_0}{2m^3}S_{a b} R_{c d e}^{a}p^b S^{cf}S_f^d p^e\\&\quad-\frac{m_0}{2m^3}S_{a b} R_{c d e}^{a}p^b S^{cf}S_f^d p^e\\&\quad-\frac{m_0}{2m^3}S_{a b} R_{c d e}^{a}p^b S^{cf}S_f^d p^e\\&\quad+\frac{m_0}{2m^3}S_{a b} R_{c d e}^{a}p^b S^{cf}S_f^d p^e.\label{appe1}
\end{aligned}\end{equation}
Applying the Tulczyjew-Dixon SCC Eq.  (\ref{sscc}), the first three terms of the right hand side of Eq. (\ref{appe1}) vanish. After defining $E_{ab}\equiv R_{acbd}p^c p^d$, we can know that $E_{ab}$ is symmetric. Besides, we have $S_{a d} S^{d e} S_{e b}$ being anti-symmetric. As a result, the last two terms of the right hand side of Eq. (\ref{appe1}) become zero. Lastly, we have  $\mathrm{D}S/\mathrm{D}\tau=0$, which is consistent with the result shown in Ref. \cite{Steinhoff:2012rw}.

\section{Derivation of Eq. (\ref{eq30})}\label{uvqr}
As suggested by \cite{Steinhoff:2012rw}, we can contract Eq. (\ref{mpd1}) with $p_\nu$ and contract Eq. (\ref{mpd2}) with $S^{\mu\nu}$, then according to the Tulczyjew SSC condition, we have 
\begin{equation}
\begin{aligned}
0=&\frac{{\rm{D}}(S^{ab} p_{b})}{{\rm{D}}\tau}=-m_0 p^a+m u^a+\frac{1}{2}R_{bcde}S^{ac}S^{de}p^b\\&-\frac{m_0 p^b R^a_{dbe}S^e_c S^{cd}}{m^2}-\frac{m_0 p^a p^b p^c R_{becf}S_d^f S^{de}}{m^3}\\&-\frac{m_0 p^b p^c p^d R^a_{fbg}R_{cidj}S_e^g S^{ef}S_h^j S^{hi}}{2m^4}\\=&-m_0 p^a+m u^a+\frac{1}{2}R_{bcde}S^{ac}S^{de}p^b\\&-\frac{m_0 p^b R^a_{dbe}S^e_c S^{cd}}{m^2}-\frac{m_0 p^a p^b p^c R_{becf}S_d^f S^{de}}{m^3}+\mathcal{O}(\epsilon^3),
\end{aligned}
\end{equation}
then we can have Eq. (\ref{eq30}) as $m^2=m_0^2+\mathcal{O}(\epsilon^3)$ \cite{Steinhoff:2012rw}.

\end{document}